\documentclass[conference]{IEEEtran}
\makeatletter
\def\ps@headings{%
\def\@oddhead{\mbox{}\scriptsize\rightmark \hfil \thepage}%
\def\@evenhead{\scriptsize\thepage \hfil \leftmark\mbox{}}%
\def\@oddfoot{}%
\def\@evenfoot{}}
\makeatother
\usepackage[dvips]{graphicx}
\usepackage{graphics}
\usepackage{graphicx}
\usepackage{subfigure}
\usepackage{amssymb}
\usepackage{amsmath}
\usepackage{amsfonts}         
\usepackage{color}            
\usepackage{epic}             
\usepackage{eepic}            
\usepackage{rotating}         
\usepackage{type1cm}          
\usepackage{epsfig}
\usepackage{amsthm}
\usepackage{epsfig}
\usepackage{latexsym}
\usepackage{amsfonts}
\usepackage{amssymb}
\usepackage{xspace}
\usepackage{mathrsfs}
\usepackage{psfrag}
\usepackage{cite}
\usepackage[ruled]{algorithm}
\usepackage[noend]{algorithmic}
\usepackage{paralist}

\newtheorem{theorem}{Theorem}[section]

\newtheorem{definition}{Definition}

\def\attauth{{\mathcal{A}}}

\begin{document}

\title{Privacy Preserving Cloud Data Access With Multi-Authorities}

\author{
Taeho Jung$^\mathcal{x}$, Xiang-Yang Li$^\mathcal{x}$, Zhiguo Wan$^\mathcal{y}$ and Meng Wan $^\mathcal{z}$\\
$^\mathcal{x}$Department of Computer Science, Illinois Institute of Technology, Chicago, IL\\
$^\mathcal{y}$School of Software, TNLIST, Tsinghua University, Beijing\\
$^\mathcal{z}$Center for Science and Technology Development, Ministry of Education, Beijing
}

\maketitle

\begin{abstract}
Cloud computing is a revolutionary computing paradigm which enables flexible, on-demand and low-cost usage of computing resources. Those advantages, ironically, are the causes of security and privacy problems, which emerge because the data owned by different users are stored in some cloud servers instead of under their own control. To deal with security problems, various schemes based on the Attribute-Based Encryption have been proposed recently. However, the privacy problem of cloud computing is yet to be solved. This paper presents an anonymous privilege control scheme \emph{AnonyControl} to address not only the data privacy problem in a cloud storage, but also the user identity privacy issues in existing access control schemes. By using multiple authorities in cloud computing system, our proposed scheme achieves anonymous cloud data access and fine-grained privilege control. Our security proof and performance analysis shows that \emph{AnonyControl} is both secure and efficient for cloud computing environment.
\end{abstract}

\footnotetext[1]{The research of authors is partially supported by NSFC under Grant No. 61170216, No.61170220, No. 61228202, No. 61003223 and No. 61128005, China 973 Program under Grant No.2011CB302705, NSF CNS-0832120, NSF CNS-1035894, NSF ECCS-1247944.}

\section{Introduction}\label{intro}
Cloud computing is a new concept of computing technique, by which
computer resources are provided dynamically via Internet. 
It attracts considerable attention and interest from both
academia and industry. However, it also has at least three challenges
that must be handled before applied to our real life. First of all,
data confidentiality should be guaranteed. When sensitive information
is stored in cloud servers, which is out of users' control in most
cases, risks would rise dramatically. The servers might illegally
inspect users' data and access sensitive information. On the other
hand, unauthorized users may also be able to intercept someone's data
(e.g. server compromise). Secondly, personal information (defined by a
user's attributes) is at risk because one's identity is authenticated
according to his information. As people are becoming more concerned
about their privacy these days, the privacy-preservability is very
important. Preferably, any authority or server alone should not know
any client's personal information. Last but not least, the cloud computing
system should be resilient in the case of security breach in which
some part of the system is compromised by attackers.

In fact, various techniques have been proposed and/or used  to address the
aforementioned problems. Identity-based encryption (IBE) was first
introduced by Shamir in 1985 \cite{shamir1985identity}. In the IBE,
the sender of a message can 
specify an identity such that only a receiver with matching identity
can decrypt it. This is different from Public-key Encryption, in that
the encrypter does not need to issue extra key to decrypter for each
ciphertext. In the IBE, the private key, which contains the identity
of the holder, is distributed to every user only once when he joins
the systtem.  

Few years later, Sahai and Waters proposed a new type of IBE -- Fuzzy
Identity-Based Encryption \cite{abe}, which is also known as
Attribute-Based Encryption(ABE). In their work, an identity is viewed
as a set of descriptive attributes. Different from the IBE, where the
decrypter could decrypt the message if and only if his identity is
exactly the same as what specified by the encrypter, this fuzzy IBE
enables the decryption if there are `identity overlaps' exceeding a
pre-set threshold between the one specified by encrypter and the one
belongs to decrypter. However, this kind of threshold-based scheme was
limited for designing more general system because the threshold based
semantic cannot express a general condition. 

Before long, more general tree-based ABE schemes, Key-Policy
Attribute-Based Encryption (KP-ABE) \cite{kp-abe} and
Ciphertext-Policy Attribute-Based Encryption (CP-ABE) \cite{cp-abe},
are proposed by Goyal \textit{et al.} and Bethencourt \textit{et al.}
respectively to overcome the aforementioned drawback of fuzzy
IBE. They look similar, but ciphertext and key structures are totally
different, and the decision of encryption policy (who can or cannot decrypt the message) is made by different parties. 

In the KP-ABE \cite{kp-abe}, a ciphertext is associated with a set of
attributes, which partially represents the ciphertext's encryption
policy. A private key is associated with a monotonic access structure
like a tree, which describes this user's identity (e.g. IIT AND (Ph.D
OR Master)). A user can decrypt the ciphertext if and only if the
access tree in his private key is satisfied by the attributes in the
ciphertext. However, the encryption policy is described in the keys,
so the encrypter does not have entire control over the encryption
policy (who has access to the data and who does not). He has to trust
that the key generators issue correct keys to correct
users. Furthermore, when a re-encryption occurs, all of the users in
the same system must have their private keys re-issued so as to gain
access to the re-encrypted files, and this process causes considerable
problems in implementation. On the other hand, those problems and
overhead are all solved in the CP-ABE \cite{cp-abe}. In the CP-ABE,
ciphertexts are created with an access structure, which specifies the
encryption policy, and private keys are generated according to users'
attributes. A user can decrypt the ciphertext if and only if his
attributes in the private key satisfy the access tree specified in the
ciphertext. By doing so, the encrypter holds the ultimate authority
about the encryption policy. Also, the already issued private keys
will never be modified unless the whole system crashes and the system's master key is lost. 

In \cite{multi1,multi2}, Chase introduced a multi-authority
system, where each user has an ID and they can interact with each key
generator (authority) using different pseudonyms. One user's different
pseudonyms are tied to his private key, but key generators never know
about the private keys, and thus they are not able to link multiple
pseudonyms belonging to the same user. In fact they are even not able
to distinguish the same user in different transactions. Also, the
whole attributes set is divided into $N$ disjoint sets and managed by
$N$ attributes authorities. That is, an attribute authority will only
issue key components which it is in charge of. In this setting, even
if an authority successfully guesses a user's ID, it knows only parts
of the user's attributes, which are not enough to figure out the
user's identity. However, the scheme proposed by Chase \emph{et
  al.}\cite{multi2} considered the basic threshold-based ABE, which is
mentioned at the beginning of this section, and thus lacks
expressibility in terms of encryption policy. 

In addition, many similar literature works
\cite{hibe,efficient-revocation,hidden-tree,ostrovsky2007abe,lewko2011decentralizing}
have been published to create more advanced schemes where data needs to
be securely and efficiently protected, which in turn served as the
base of the research on security protocol in cloud computing
environment
\cite{kp-abe-cloud,habe,hasbe,revoke-cloud}. 
However, much less effort is paid to protect
users' privacy during those interactive protocols. Users' identities,
which are described with their attributes, are generally opened to key
generators, and the generators issue private keys according to their
attributes. But it seems natural that users might want to keep their
identities secret while they still get their private keys. Lewko's work \cite{lewko2011decentralizing} 
is the most similar one to ours in that they also tried to decentralize 
the central authority in the CP-ABE into multiple ones. They use a LSSS matrix as an access structure, but their scheme only converts the AND, OR gates to the LSSS matrix while we inherit the flexibility of the access tree having threshold gates. In addition, our system tolerates the compromise attack towards attributes authorities, which is not covered in many existing works.

The main contributions of this paper are: 
\begin{compactenum}
\item 
The proposed scheme is able to protect
user's privacy against each single authority.
 \item 
The proposed scheme
is tolerant against authority compromise, and compromising of up to
$(N-2)$ authorities does not bring the whole system down.
\item 
We provide detailed
analysis on security and performance to
show feasibility of our scheme.
\item
We first implement the real toolkit of multi-authority based encryption scheme.
\end{compactenum}

\section{Preliminaries \& Related Work}
\subsection{Preliminaries}

Let $\mathbb{G}_0$ be a multiplicative cyclic group of prime order $p$ and $g$ be its generator. The bilinear map $e$ is defined as follows:  $e:\mathbb{G}_0\times\mathbb{G}_0\rightarrow\mathbb{G}_T$, where $\mathbb{G}_T$ is the codomain of $e$. The bilinear map $e$ has the following properties:
\begin{compactitem}
\item
\emph{Bilinearity}:  $\forall u,v\in \mathbb{G}_0$ and $a,b\in \mathbb{Z}_p$, $e(u^a,v^b)=e(u,v)^{ab}$.
\item
\emph{Symmetry}: for all $u,v\in \mathbb{G}_0$, $e(u,v)=e(v,u)$.
\item
\emph{Non-degeneracy}: $e(g,g)\neq 1$.
\end{compactitem}

\begin{definition}
\label{def:DDH}
The Decisional Diffie-Hellman (DDH) problem in group $\mathbb{G}_0$ of
prime order $p$ with generator $g$ is defined as follows: on input $g,
g^a, g^b, g^c=g^{ab}\in\mathbb{G}_0$, where $a,b,c\in \mathbb{Z}_p$,
decide whether $c=ab$ or $c$ is a random element.
\end{definition}

\begin{definition}
\label{def:DBDH}
The Decisional Bilinear Diffie-Hellman (DBDH) problem in group
$\mathbb{G}_0$ of prime order $p$ with generator $g$ is defined as
follows: on input $g, g^a, g^b, g^c\in\mathbb{G}_0$ and
$e(g,g)^{z}=e(g,g)^{abc}\in\mathbb{G}_T$, where $a,b,c\in
\mathbb{Z}_p$, decide whether $z=abc$ or $z$ is a random element.
\end{definition}

The security of many ABE schemes
(e.g. \cite{cp-abe,hasbe,hidden-credential,hidden-policy})
and ours rely on the assumption that no probabilistic polynomial-time
algorithms can solve the DDH and DBDH problem with non-negligible
advantage. This assumption is reasonable since discrete logarithm
problems in large number field are widely considered to be
intractable (\cite{jung2013data,li2013search,zhang2013verifiable,zhang2012message}), and therefore $a$ is not deducible from $g^a$ even if $g$ is publicly known.

We also define the Lagrange coefficient $\triangle_{i,S}$ for
$i\in\mathbb{Z}_p$ and a set, $S$, of elements in $\mathbb{Z}_p$: $\triangle_{i,S}(x):=\prod_{j\in S,j\neq i}\frac{x-j}{i-j}$
, which will be used in the polynomial
interpolation in decryption algorithm to recover the secretly shared values. Additionally, a one-way hash function
$H:\{0,1\}^*\rightarrow \mathbb{G}_0$ is also defined as a random
oracle. This will map any attribute value to a random element in
$\mathbb{Z}_p$ (same attribute value will be mapped to the same element).

\subsection{CP-ABE}
The idea of Attribute-Based Encryption (ABE) was first proposed by
Sahai and Waters \cite{abe}. On the contrary to the traditional
identity-based encryption, a user is able to decrypt a ciphertext if
there is some match between his private key and ciphertext in the
ABE. However, due to its lack of expressibility and generalization, it
was later extended to the Key-Policy ABE by Goyal \emph{et al.}
\cite{kp-abe} and the Ciphertext-Policy ABE by Bethencourt \emph{et
  al.} \cite{cp-abe}. Our scheme chooses CP-ABE as the base due to its
advantages mentioned in the Section \ref{intro}.

In the CP-ABE, the private key is distributed to users by a trusted
central issuer only once. The keys are identified with a set of
descriptive attributes, and the encrypter specifies an encryption
policy using an access tree so that those with private keys which
satisfy it can decrypt the ciphertext.


\subsection{Privilege Trees $T_p$}
In most of previous works \cite{kp-abe,cp-abe,hasbe},
encryption policy is described with a tree called access tree. Each
non-leaf node of the tree is a threshold gate, and each leaf node is
described by an attribute. One access tree is required in every data
file to define the encryption policy. In this paper, we extend existing schemes by generalizing the access tree to a
privilege tree. The privilege in our scheme is defined as
follows. A data file has several operations executable on itself, but some of them should be restricted only to
authorized users. For example, \{\emph{Read\_mine, Read\_all, Delete,
Modify, Create}\} is a privileges set of students' grades. Then, reading Alice's 
grades is allowed to her and her professors, but all other privileges should be
authorized only to the professors, so we need to grant the ``\emph{Read\_mine}'' 
to Alice and all other to the professors.

Every operation is associated with one privilege $p$,
 which is described by a privilege tree $T_p$.
If a user's attributes satisfy $T_p$, he is granted
the privilege $p$. By doing so, we not only control the file access
but also control other executable operations, which makes the file
controlling fine-grained and thus suitable for cloud storage service.

In our scheme, several trees are required in every data file to verify
users' identity and to grant him a privilege accordingly. There are
supposed to be $r$ these kind of structures, which means there are $r$
different privileges defined for the corresponding data file. The
privilege 0 is defined as the privilege to read the file, and other privileges may be defined arbitrarily (the
$m^{th}$ privilege does not necessarily have more powerful privilege than the $n^{th}$ one
when $m>n$). The tree is similar to the one defined in
\cite{cp-abe}. Given a tree, if $\emph{num}_x$ is the number of the
node \emph{x}'s children node and $\emph{k}_x$ is its threshold value
$0<\emph{k}_x\leq\emph{num}_x$, then node $x$ is assigned a true value
if at least $k_x$ children nodes have been assigned true
value. Specially, the node becomes an OR gate when $\emph{k}_x=1$ and
an AND gate when $\emph{k}_x = \emph{num}_x$.

Several subsidiary functions are to be mentioned for convenience. We denote the parent of a node $x$ by $parent(x)$, and the attribute value of a leaf node $x$ by $att(x)$. Furthermore, the privilege tree $T_p$ also defines the order between children of every node, and the numbers associated with node $x$, from 1 to $num_x$, are denoted by $index(x)$.\vspace{-5pt}
\begin{table}[!ht]
\centering 
\caption{Notations for privilege trees}
\vspace{-3pt}
\label{table:notation_tree}
\centering
\begin{tabular}{l|l}
\hline \hline
$T_p$ & $p^{th}$ privilege tree representing the $p^{th}$ privilege\\
$k_x$ & threshold value of the node $x$ \\
$num_x$ & number of $x$'s child nodes \\
$att(x)$ & attribute value of the node $x$, if it is a leaf node \\
$index(x)$ & index of the $x$'s child nodes \\
$parent(x)$ & node $x$'s parent node \\
\hline \hline
\end{tabular}
\end{table}
\vspace{-5pt}

\subsection{Satisfying the Privilege Tree}

If a user's attributes set $S$ satisfies the privilege tree $T_p$ or the node $x$, we define it as $T_p(S)=1$ or $x(S)=1$ respectively. $T_p(S)$ is calculated recursively as follows. If $x$ is a leaf node, $x(S)=1$ if and only if $att(x)\in S$. If $x$ is a non-leaf node, $x(S)=1$ only when at least $k_x$ child nodes return 1. For the root node $R_p$ of $T_p$, $T_p(S)=1$ only if $R_p(S)=1$.

\begin{figure}[!h]
\begin{center}
\scalebox{0.45}{\input{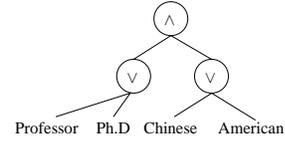}}
\vspace{-5pt}
\caption{An example of a privilege tree.}
\vspace{-17pt}
\label{fig:tree}
\end{center}
\end{figure}
Figure \ref{fig:tree} shows an example of the privilege tree $T_p$ for deleting the file. For an instance, if a user's attributes set is \{Sex:Male, Age:23, Nationality:Chinese, University:Tsinghua University, Position:Ph.D Student, Religion:None\}, he satisfies the tree, and thus achieving the corresponding privilege (here it means he can delete the file).

\section{Definitions of our scheme}
\label{sec:definition}

\subsection{System Model}
In our system, there are four types of entities: $N$ \emph{Attribute Authorities} (denoted as $\attauth$), \emph{Cloud Server}, \emph{Data Owners} and \emph{Data Consumers}. A user can be a Data Owner and a Data Consumer simultaneously.

\begin{figure}[!th]
  \centering
  \includegraphics[width=0.4\textwidth]{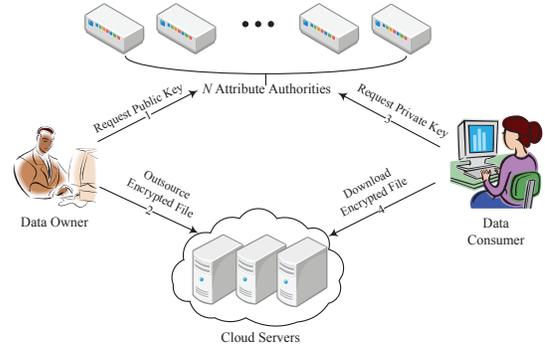}
  \vspace{-5pt}
  \caption{Our system model}
  \vspace{-5pt}
  \label{fig:model}
\end{figure}
Authorities are assumed to have powerful computation abilities, who are supervised by government offices since keys act as IDs and partially contain users' PII (Personally Identifiable Information). The whole attribute set is divided into $N$ disjoint sets and controlled by each authority. One practical method to divide the attributes set is to divide them by category (e.g., \{\textbf{Sex}: Male, Female\}, \{\textbf{Nationality}: Korean, Chinese, Japanese\}, \{\textbf{University}: Tsinghua, Peking University\}, \{\textbf{Position}: Professor, Ph.D Student, Master Student\}). In this way, since each authority is aware of only one type of attribute, no useful information is leaked. The authorities jointly compute a system-wide public key, and individually compute their master keys at the initialization phase. The public key is used for all operations within the system, and the master keys are used by each attribute authority when he generates private keys for Data Consumers.

A Data Owner achieves public key from any one of the authorities, and he uses the public key to encrypt the data file before outsourcing it to the Cloud Servers. The Cloud Server, who is assumed to have adequate storage capacity, does nothing but store them.

Newly joined Data Consumers request private keys from all of the authorities, and they do not know which attributes are controlled by the authorities. On the other hand, authorities do not know which Data Consumers are interacting with them because each of them knows only a part of Data Consumers' attributes. When the Data Consumers request their private keys from the authorities, authorities jointly create corresponding private key and send it to them.

All Data Consumers are able to download any of those data files, but only those whose private keys satisfy the privilege tree $T_p$ can execute the operation associated with privilege $p$. When a user wants to execute a specific operation upon a data, he should satisfy the relevant privilege tree $T_p$ and gets verified by the Cloud Server. The server is delegated to execute an operation $p$ if and only if the user's privilege is verified through the privilege tree $T_p$.

\subsection{Threats Model}

We assume the Cloud Servers are untrusted, who behave properly in most of time but may collude with malicious Data Consumers or Data Owners to harvest others' file contents to gain illegal profits. But they are also assumed to gain legal benefit when users' requests are correctly processed, which means they will follow the protocol in general. In addition, even if the Cloud Server illegally modifies data files for sake of monetary benefits (e.g. deleting rarely accessed files to save the storage), whether the data is intact can be detected by the TPA technique introduced in \cite{public-audit}.

The $N$ authorities are assumed to be semi-honest. That is, they will follow our proposed protocol in general, but try to find out as much information as possible individually. More specifically, we assume they are interested in users' attributes to achieve the identities, but they will never collude with any user or authority to harvest file contents even if it is highly beneficial. This assumption is similar to many previous researches on security issue in cloud computing (e.g. \cite{hasbe,public-audit,linear-outsourcing,rank-search}), and it is also reasonable since these authorities will be audited by government offices.

Data Consumers are untrustful since they are random users including attackers. They may collude with other Data Consumers to access what not allowed for them.

\subsection{Design Goal}

Our goal is to help Data Owners securely share their data with Data Consumers, where fine-grained privilege control is achievable, and to guarantee the confidentiality of Data Consumers' identity information by decomposing a center authority to multiple ones while preserving tolerance to compromise attacks on the authorities. We assume the identity information is not disclosed by the underlying network. This can be achieved by employing anonymized protocols (e.g., \cite{liu2011rumor}).

In the rest of this paper, $\mathbb{A}^u$ is used to denote the attributes set of a user $u$. $\mathcal{A}_k$ is used to denote the attribute authority $k$, and we also use a subscript $k$ to denote the attributes set handled by $\mathcal{A}_k$.

\begin{table}[!ht]
\centering 
\vspace{-5pt}
\caption{Notations for System Construction}
\vspace{-7pt}
\label{table:notation_system}
\centering
\begin{tabular}{l|l}
\hline
\hline
Attribute Authorities\\
\hline
$k$ & index of an attribute authority \\
$\mathcal{A}_k$ & the $k^{th}$ attribute authority \\
$s_{kj}$ & secret parameter for compromise tolerance \\
\hline
\hline
Data Owner\\
\hline
$u$ & a user (either Data Owner or Data Consumer)\\
$\mathbb{A}^u$ & attributes set of user $u$ \\
$\mathbb{A}^{T_p}$ & attributes set included in tree $T_p$ \\
$K_e$ & symmetric encryption/decryption key \\
\hline
\hline
\end{tabular}
\end{table}
\vspace{-14pt}

\subsection{Definition of our Multi-Authority CP-ABE}
\noindent\textsf{Setup $\rightarrow \mathbf{PK}, \mathbf{MK}_k$:}

This algorithm takes nothing as input. Attributes authorities execute this algorithm to jointly compute a system-wide public parameter $\mathbf{PK}$, authority-wide public parameter $y_k$ and to individually compute master keys $\mathbf{MK}_k$.

\noindent\textsf{KeyGenerate($\mathbf{PK}$, $\mathbf{MK_k}$, $\mathbb{A}^u$, $\mathbf{GID}_u$) $\rightarrow \mathbf{SK}_u$:}

This algorithm enables a user to interact with every attribute authority, and obtains a private key $\mathbf{SK}_u$ corresponding to the input attribute set $\mathbb{A}^u$ and his global ID $\mathbf{GID}_u$. From the authorities' perspective, the algorithm enables them to jointly generate a private key $\mathbf{SK}_u$, using the public key $\mathbf{PK}$ and master keys $\mathbf{MK}_k$, according to the input $\mathbb{A}^u$ and a pseudonym $nym_{\mathbf{GID}_u}$, which is created according to the $\mathbf{GID}_u$. Authorities are not able to derive a user's $\mathbf{GID}_u$ based on the pseudonym $nym_{\mathbf{GID}_u}$.

\noindent\textsf{Encrypt($\mathbf{PK}$, $M$, $\{T_p\}_{p\in\{0,\cdots,r-1\}}$) $\rightarrow (\mathbf{CT}, \mathbf{VR}$):}

This algorithm takes as input the public key $\mathbf{PK}$, a message $M$, and a set of privilege trees $\{T_p\}_{p\in\{0,\cdots,r-1\}}$, where $r$ is determined by the encrypter. It will encrypt the message $M$ and returns a ciphertext $\mathbf{CT}$ and a verification set $\mathbf{VR}$ so that a user can execute specific operation on the ciphertext if and only if his attributes satisfy the corresponding privilege tree $T_p$. As we defined, $T_0$ stands for the privilege to read the file.

\noindent\textsf{Decrypt($\mathbf{PK}$, $\mathbf{SK}_u$, $\mathbf{CT}$) $\rightarrow M$ or verification parameter:}

This algorithm will be used at file controlling (e.g. reading, modification, deletion). It takes as input the public key $\mathbf{PK}$, a ciphertext $\mathbf{CT}$, and a private key $\mathbf{SK}_u$, which has a set of attributes $\mathbb{A}^u$ and corresponds to its holder's $\mathbf{GID}_u$. If the set $\mathbb{A}^u$ satisfies any tree in the set $\{T_p\}_{p\in\{0,\cdots,r-1\}}$, the algorithm returns a message $M$ or a verification parameter. If the verification parameter is successfully verified by Cloud Servers, who use $\mathbf{VR}$ to verify it, the operation request will be processed.

\noindent\textsf{ReEncrypt($\mathbf{PK}$, $\mathbf{CT}$, $\mathbf{SK}_u$, $\{T'_p\}$) $\rightarrow \mathbf{CT}', \mathbf{VR}'$:}

This algorithm is barely a composition of the decryption and the encryption algorithm. It takes as input the public key $\mathbf{PK}$, a private key $\mathbf{SK}_u$, a ciphertext $\mathbf{CT}$, and a set of new privilege trees $\{T'_p\}_{p\in\{0,\cdots,r-1\}}$.
If the set $\mathbb{A}^u$ in $\textbf{SK}_u$ satisfies $T_0$ (to obtain the original message $M$ first) and $T_k$ (privilege $k$ is the one for re-encryption), the algorithm re-encrypts the original message $M$ under new set of privilege trees and returns a new ciphertext $\mathbf{CT}'$ and a new verification set $\mathbf{VR}'$

\section{Our AnonyControl Scheme}
\subsection{Setup}
At the system initialization phase, any one of the authorities chooses a bilinear group $\mathbb{G}_0$ of prime order $p$ with generator $g$ and publishes it. Then, all authorities independently and randomly picks $v_k\in \mathbb{Z}_p$ and send $Y_k=e(g,g)^{v_k}$ to all other authorities who individually compute $Y:=\prod_{k \in {\attauth}} Y_k=e(g,g)^{\sum_{k \in {\attauth}} v_k}$.

Then, every authority $\mathcal{A}_k$ randomly picks $N-1$ integers $s_{kj}\in\mathbb{Z}_p (j\in \{1,\cdots,N\}\backslash\{k\})$ and computes $g^{s_{kj}}$. Each $g^{s_{kj}}$ is shared with each other authority $\mathcal{A}_j$ An authority $\mathcal{A}_k$, after receiving
$N-1$ pieces of $g^{s_{jk}}$ generated by $\mathcal{A}_j$, computes its secret parameter $x_k \in \mathbb{Z}_p$ as follows:

\begin{displaymath}
\begin{split}
x_k &=(\prod\limits_{j\in \{1,\cdots,N\}\backslash\{k\}}^{}{g^{s_{kj}}}) / (\prod\limits_{j\in \{1,\cdots,N\}\backslash\{k\}}^{}{g^{s_{jk}}})\\
&=g^{(\sum\limits_{j\in \{1,\cdots,N\}\backslash\{k\}}^{} {s_{kj}} - \sum\limits_{j\in \{1,\cdots,N\}\backslash\{k\}}^{} {s_{jk}})}
\end{split}
\end{displaymath}

It is easy to see that these randomly produced integers satisfy $\prod_{k \in \attauth}{x_k}=1 \mod p$. This is an important property which achieves compromise attack tolerance for our scheme, which will be discussed in the next section.

Then, the master key for the authority $\mathcal{A}_k$ is $\mathbf{MK}_k=\{v_k,x_k\}$, and public key of the whole system is published as $\mathbf{PK}=\{\mathbb{G}_0, g, Y=e(g,g)^{\sum v_k}\}$.

Note that the time complexity of the setup computation is $O(N^2)$ since every authority computes $N-1$ pieces of $g^{s_{kj}}$. However, this can be further reduced to $O(N)$ by applying the following simple trick. We first cluster the authorities into $C$ clusters, and exchanges the parameters within the cluster only. Then, the time complexity is reduced to $O(CN)=O(N)$ since $C$ is a constant. Yet, this trick will also relax our compromise tolerance also. We will further discuss the trade-off between time complexity and compromise attack tolerance in the Section \ref{sec:securiy_analysis}.


\subsection{KeyGenerate($\mathbf{PK}$, $\mathbf{MK_k}$, $\mathbb{A}^u$)}
When a new user $u$ with $\mathbf{GID}_u$ wants to join the system, he requests the private key from all of the authorities by following this process which is composed of two phases.

\subsubsection{Attribute Key Generation}
For any attribute $i\in \mathbb{A}^u$, every $\mathcal{A}_k$ randomly picks $r_i\in\mathbb{Z}_p$ to individually compute the partial private key
\begin{displaymath}
H(att(i))^{r_i},D_i'=g^{r_i}
\end{displaymath}

Then, all of the authorities randomly picks $d_k\in \mathbb{Z}_p$ ($d_k$ for $\mathcal{A}_k$) and compute $x_k\cdot g^{v_k}\cdot g^{d_k}$ and share it with others.

Then, authorities merge the partial private keys by computing the following: (this can be individually done by any one of the authorities)
\begin{displaymath}
\begin{split}
D&=\prod{x_k g^{v_k} g^{d_k}}=g^{\sum{v_k}+\sum{d_k}}\\
D_i&=H(att(i))^{r_i}\cdot \prod{g^{d_k}}
=H(att(i))^{r_i}\cdot g^{(\sum {d_k})}
\end{split}
\end{displaymath}
At the end, the $D$, $D_i$'s and $D_i'$'s are sent to the user $u$.

\subsubsection{Key Aggregation}
User $u$, after receiving $D_i$'s and $D_i'$'s, aggregates the components as his private key:
\begin{displaymath}
\begin{split}
\mathbf{SK}_u=\{D,\forall i\in\mathbb{A}^u:D_i=g^{(\sum d_k)}\cdot H(att(i))^{r_i},D_i'=g^{r_i}\}
\end{split}
\end{displaymath}

\subsection{Encrypt($\mathbf{PK}$, $M$, $\{T_p\}_{p\in\{0,\cdots,r-1\}}$)}
Encryption must be done before Data Owners upload their data files to the Cloud Server. At first, he randomly selects a symmetric data encryption key $K_e$ and encrypts the data file with it using any symmetric encryption scheme (e.g., 256-bit AES). Then, he determines a set of privilege trees $\lbrace T_p\rbrace_{p\in\lbrace 0,\cdots,r-1\rbrace}$ and executes $Encrypt(\mathbf{PK},K_e,\lbrace T_p\rbrace)$. For each $T_p$, the algorithm first chooses a polynomial $q_x$ for each node $x$ in it. For each node $x$, sets the degree $d_x$ of the polynomial $q_x$ as one less than the threshold value $k_x$. Starting from the root node $R_p$, the algorithm randomly picks $s_p\in\mathbb{Z}_p$ and sets $q_{R_p}(0):=s_p$. Then, it chooses other numbers so that for any other node $x$, $q_x(0)=q_{parent(x)}(index(x))$ and randomly defines the $q_x$ too.

Here, Shamir's secret sharing technique \cite{shamir1979share} is directly used to implement the threshold gate. Shamir's $t$-out of-$n$ secret share scheme allows one to divide a secret to $n$ shares, and the original secret can be recovered with $t$ of them. So, in our tree, the node value of the gate is recovered if and only if at least $k_x$ values of children nodes are recovered in recursive manner.

Finally, the ciphertext $\mathbf{CT}$ is created as
\begin{displaymath}
\begin{split}
\mathbf{CT}=\langle &\lbrace T_p\rbrace_{p\in\lbrace 0,\cdots,r-1\rbrace },E_0=K_e\cdot Y^{s_0},\\
&\{C_i=g^{q_i(0)},C_i'=H(att(i))^{q_i(0)}\}_{i\in \mathbb{A}^{T_p},\forall p\in\{0,\cdots,r-1\}} \rangle
\end{split}
\end{displaymath}

In the $\mathbf{CT}$ above, $E_0$ contains the symmetric key for decryption, and $C_i$'s and $C'_i$'s represent the attribute values in the specified privilege trees.

Then, $\mathbf{VR}$, which is disclosed only to the Cloud Server, is created for the purpose of privilege verification.
\begin{displaymath}
\begin{split}
\mathbf{VR}=\langle\lbrace E_p=Y^{s_p}\rbrace_{p\in\lbrace 1,\cdots,r-1\rbrace}\rangle
\end{split}
\end{displaymath}

Finally, Data Owner selects a unique $ID$ for this encrypted data file and sends $\mathbf{CT}$, $\mathbf{VR}$ and the encrypted file to the Cloud Server to share them with other Data Consumers.
\begin{figure}[!h]
  \centering
\includegraphics[width=0.25\textwidth]{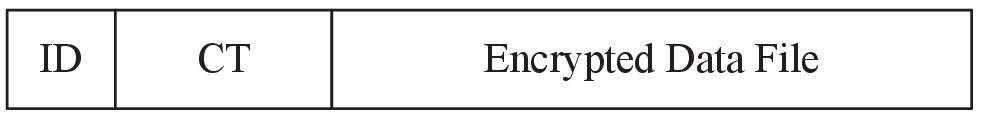}\\
(a) {\small {seen from the data consumers' perspective}}.\\
 \includegraphics[width=0.3\textwidth]{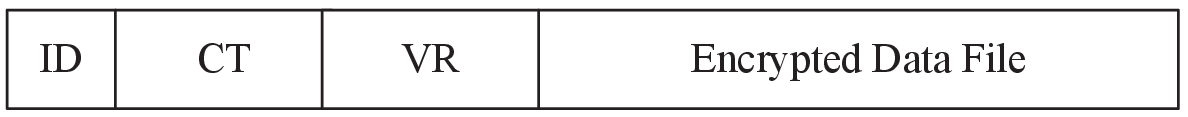}\\
(b) {\small {seen from the server's perspective.}}
  \caption{ A data file stored on the cloud.}
  \label{fig:file}
  \label{fig:file_server}
\end{figure}
\vspace{-5pt}

\subsection{Decrypt($\mathbf{PK}$, $\mathbf{SK}_u$, $\mathbf{CT}$)}
Every user within the system can download the ciphertext from the Cloud Server, but he is able to execute operations upon encrypted data only after he successfully decrypts it. Firstly, we define a recursive algorithm $DecryptNode(\mathbf{CT},\mathbf{SK}_u,x)$, where $x$ stands for a node in the privilege tree $T_p$. If the node $x$ is a leaf node, we let $i$ be the attribute of the node $x$ and define as follows. If $i \in \mathbb{A}^u$,

\begin{eqnarray*}
&DecryptNode(\mathbf{CT},\mathbf{SK}_u,x)=\frac{e(D_i,C_x)}{e(D_i',C_x')}\\
=&\frac{e(g^{\sum d_k}\cdot H(att(i))^{r_i},g^{q_x(0)})}{e(g^{r_i},H(att(i))^{q_x(0)})}
=e(g,g)^{(\sum d_k)\cdot q_x(0)}
\end{eqnarray*}
If not, we define \mbox{$DecryptNode(\mathbf{CT},\mathbf{SK}_u,x):=\bot$.}

If $x$ is not a leaf node, the algorithm proceeds as follows: For all nodes $z$ that are children of $x$, it calls $DecryptNode(\mathbf{CT},\mathbf{SK}_u,z)$ and stores the output as $F_z$. Let $S_x$ be an arbitrary $k_x$-sized set of child nodes $z$ such that $F_z\neq \varnothing$. If no such set exists then the node was not satisfied and the algorithm returns $\bot$. Otherwise, compute
\begin{displaymath}
\begin{split}
F_x&=\prod_{z\in S_z}F_z^{\triangle_{d,s_x'}(0)}, \textrm{  where}
\left\{ \begin{array}{ll}
d=index(z) & \\
S_x'={index(z):z\in S_x} &\\
\end{array} \right.\\
&=\prod_{z\in S_z}(e(g,g)^{(\sum d_k)\cdot q_z(0)})^{\triangle_{d,S_x'}(0)}\\
&=\prod_{z\in S_z}(e(g,g)^{(\sum d_k)\cdot q_{parent(z)}(d)})^{\triangle_{d,S_x'}(0)}\\
&=\prod_{z\in S_z}(e(g,g)^{(\sum d_k)\cdot q_x(d)})^{\triangle_{d,S_x'}(0)}\\
&=e(g,g)^{(\sum d_k)\cdot q_x(0)} \textrm{  (using polynomial interpolation)}\\
\end{split}
\end{displaymath}

The interpolation above recovers the parent node's value by calculating coefficients of the polynomial and evaluating the $p(0)$. We direct the readers to \cite{shamir1979share} for complete calculation. A user recursively calls this algorithm, starting from the root node $R_p$ of the tree $T_p$, after downloading the file. If the tree is satisfied, which means he is granted the privilege $p$, then
\begin{displaymath}
\begin{split}
DecryptNode(\mathbf{CT},\mathbf{SK}_u,R_p)
&=e(g,g)^{{s_p}\sum d_k}
\end{split}
\end{displaymath}

Finally, if the user is trying to read the file, the symmetric encryption key $K_e$ can be recovered by:
\begin{displaymath}
\frac{E_0}{\frac{e(g,D)}{e(g,g)^{s_0\sum{d_k}}}}=\frac{K_e\cdot Y^{s_0}}{\frac{e(g,g^{\sum d_k+\sum v_k})}{e(g,g)^{\sum d_k}}}=K_e
\end{displaymath} 
Then, the data file can be decrypted by using it. Otherwise, he should be verified as an authorized user for the operation first. If the operation requires the $j^{th}$ privilege, the user recursively calls $Decrypt(\mathbf{CT},\mathbf{SK}_u,x)$ starting from the root node $R_j$ of the tree $T_j$ to get $e(g,g)^{s_j\sum d_k}$ and further achieve $Y^{s_j}$ with the same equation as above. The user sends it to the Cloud Server as well as the operation request. The Cloud Server checks whether \mbox{$Y^{s_j}=E_j$}, and proceeds if they do equal each other. In fact, $Y^{s_j}$ should be encrypted to avoid replay attack. This can be simply implemented by introducing any public key encryption protocol.

\subsection{ReEncrypt($\mathbf{PK}$, $\mathbf{CT}$, $\mathbf{SK}_u$, $\{T'_p\}_{p\in\{0,\cdots,r-1\}}$)}
In real applications in a cloud storage system, users might be revoked due to some reasons (e.g., resignation from a company). In this case, we need to re-encrypt the files to avoid unauthorized access by revoked users, the users who satisfy certain properties for revocation (e.g., resignation). When they are revoked, they should not access the data files or execute other operations on them. An authorized user with the privilege to re-encrypt the associated file (note that this user might not be limited to the Data Owner) decrypts it first, and randomly selects another symmetric encryption key $K'_e$ to re-encrypt it. Then, he determines subtrees which forbid revoked users' access but still enables other unrelated users' one, and adds these subtrees into the original $\lbrace T_p\rbrace$ to gain new privilege set $\lbrace T'_p\rbrace$. Then, $ReEncrypt(\mathbf{PK}, \mathbf{CT}, \mathbf{SK}_u, \{T'_p\})$ is executed to obtain new $\mathbf{CT}'$ and $\mathbf{VT}'$.
\begin{figure}[!th]
\begin{center}
\scalebox{0.45}{\input{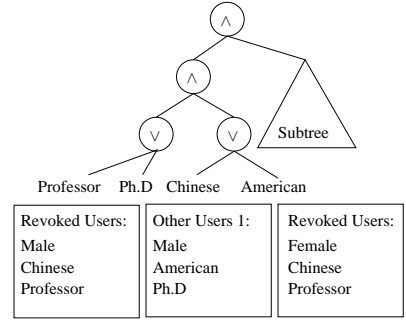}}
\vspace{-5pt}
\caption{An example of privilege tree after the re-encryption}
\vspace{-10pt}
\label{fig:tree2}
\end{center}
\end{figure}

Figure.\ref{fig:tree2} shows a new tree $T'_p$ after a re-encryption. Assuming that there are only three users within the system, who are described in Figure.\ref{fig:tree2}, and the \textquoteleft Revoked User\textquoteright\space is revoked, the subtree may indicate the attribute condition as $(\textrm{Chinese} \vee \textrm{Ph.D Student})$.



\section{Security Analysis}\label{sec:securiy_analysis}

\subsection{User's Identity Information Confidentiality}
The attributes, which contain a user's identity information, are separately controlled by different attribute authorities. Therefore, a user's attributes information is securely protected.
\subsection{Trade-off between Tolerance and Complexity}
In the proposed scheme, an authority $\mathcal{A}_k$ generates a set of random secret parameters $\{s_{kj}\}$ and shares it with other authorities, and the $x_k$ is computed based on this parameters. Even if an adversary is able to compromise up to $(N-2)$ authorities, there are still two parameters kept unknown to the adversary. So, the adversary is not able to guess the valid $g^{\sum v_k}$, and he fails to construct a valid secret key. Hence, the scheme achieves compromise tolerance to up to $(N-2)$ authorities compromise.

But, if we reduce the time complexity of the setup phase by dividing authorities into several clusters having $C$ authorities in each, attackers can compromise $C-1$ authorities in a cluster to create valid master keys of that cluster. Therefore, there is a trade-off between tolerance and complexity. However, since the number of authorities is typically not very huge, and the setup is one-time operation at the very beginning of the system setup, we recommend using the original setup algorithm whose complexity is $O(N^2)$.

Finally, note that the compromised authorities are able to issue valid attribute keys for which they are in charge of, so the ciphertexts whose privilege trees have only those attributes might be illegally decrypted if the attacker issue all possible attribute keys to himself. But, since the authorities are well protected servers, it is hard to compromise even one authority, and the probability of compromising enough authorities to illegally decrypt some ciphertext is very low.

\subsection{Data Confidentiality against Collusion Attack}
In order to access a plaintext, attackers must recover \mbox{$Y^{s_0}=e(g,g)^{s_0\sum v_k}$}, which can be recovered only if the attackers have enough attributes to satisfy the tree $T_0$. When two different keys' components are combined, the combined key cannot go through the polynomial interpolation in the decryption algorithm due to the randomization. Therefore, at least one key should be valid to satisfy the privilege tree.
\medskip

\noindent \textbf{\textsf{Security Model for \emph{AnonyControl} construction}}

W.L.O.G., we assume there is only one privilege tree $T_0$, which represents the privilege for reading, in the privilege set of $\mathbf{CT}$. This assumption does not affect the security proof since we can easily extend the model to have several trees in $\mathbf{CT}$. Next we show our scheme is secure against \textit{chosen plaintext attacks} (CPA) if all probabilistic polynomial-time adversaries have negligible advantages in this game.
\vspace{5pt}
\\
\textsf{Init} \hspace{4pt} The adversary declares a privilege tree $T_0^*$, which he wants to be challenged.
\\
\textsf{Setup} \hspace{4pt} The challenger runs the Setup algorithm of our construction and publishes $\mathbf{PK}$ to the adversary.
\\
\textsf{Phase 1} \hspace{4pt} The adversary queries for as many private keys, which correspond to attribute sets $\mathbb{A}_1,\cdots ,\mathbb{A}_q$, as he wants, where none of these keys satisfy the $T_0^*$ above.
\\
\textsf{Challenge} \hspace{4pt} The adversary submits two messages $M_0$ and $M_1$ of equal size to the challenger. The challenger flips a random binary coin $b$ and encrypts $M_b$ with $T_0^*$. The ciphertext $\mathbf{CT}^*$ is given to the adversary.
\\
\textsf{Phase 2} \hspace{4pt} Phase 1 is repeated adaptively.
\\
\textsf{Guess} \hspace{4pt} The adversary outputs a guess $b'$ of $b$.

The advantage of an adversary $\mathcal{A}$ in this game is defined as Pr$[b'=b]-\frac{1}{2}$. We note that the model can easily be extended to handle chosen-ciphertext attacks by allowing for decryption queries in Phase 1 and Phase 2.

Here we prove that the security of our scheme in the security model above reduces to the intractability of the DBDH problem.

\begin{definition}
Our scheme is secure against CPA if all polynomial-time adversaries have at most a negligible advantage in the above game.
\end{definition}
\begin{theorem}
If an adversary can break our scheme in the security model above, there exists at least one probabilistic polynomial-time algorithm can solve the DBDH problem, which is defined in the Section III, with a non-negligible advantage.
\end{theorem}
\begin{proof}
Suppose there exists a probabilistic polynomial-time adversary $\mathcal{A}$ can attack our scheme in the security model above with advantage $\epsilon$. We prove that the following DBDH game can be solved with advantage $\frac{\epsilon}{2}$.

Let $e:\mathbb{G}_0 \times \mathbb{G}_0 \rightarrow \mathbb{G}_T$ be a bilinear map, where $\mathbb{G}_0$ is a multiplicative cyclic group of prime order $p$ and $g$ is its generator. First the DBDH challenger flips a binary coin $\mu$, and he sets $(g,A,B,C,Z):=(g,g^a,g^b,g^c,e(g,g)^{abc})$ if $\mu=0$; otherwise he sets $(g,A,B,C,Z):=(g,g^a,g^b,g^c,e(g,g)^{z})$, where $a,b,c,z\in \mathbb{Z}_p$ are randomly picked.  The challenger then gives the simulator $\langle g, A, B, C, Z \rangle = \langle g, g^a, g^b, g^c, Z \rangle$. The simulator $sim$ then plays the role of a challenger in the following DBDH game.
\vspace{5pt}
\\
\textsf{Init} \hspace{4pt} The adversary $\mathcal{A}$ creates a $T_0^{*}$ which he wants to be challenged (Nodes inside the tree should be defined by him).
\\
\textsf{Setup} \hspace{4pt}
$sim$ sets the parameter $Y:=e(A,B)=e(g,g)^{ab}$ and gives this public parameter to $\mathcal{A}$.
\\
\textsf{Phase 1} \hspace{4pt}
$\mathcal{A}$ queries for as many private keys, which correspond to attribute sets $\mathbb{A}_1,\cdots ,\mathbb{A}_q$, as he wants, where none of them satisfy the $T_0^{*}$. $sim$, after receiving the key queries, computes the components in private keys to respond the $\mathcal{A}$'s requests. For all attributes $i\in \mathbb{A}^u$, he randomly picks $r_i\in \mathbb{Z}_p$, and computes $D_i:=A\cdot H(att(i))^{r_i}, D_i':=g^{r_i}$. Then, $sim$ returns the created private key to $\mathcal{A}$.
\\
\textsf{Challenge} \hspace{4pt} The adversary $\mathcal{A}$ submits two challenge messages $m_0$ and $m_1$ to the challenger. The challenger flips a binary coin $\gamma$, and returns the following ciphertext to $\mathcal{A}$.
\begin{displaymath}
\begin{split}
\mathbf{CT^*}=\langle &T_0^{*}, E_0=m_{\gamma}\cdot Z,\\
&\{C_i=g^{q_i(0)},C_i'=H(att(i))^{q_i(0)}\}_{i\in \mathbb{A}^{T_0^{*}}} \rangle
\end{split}
\end{displaymath}

If $\mu=0$, $Z=e(g,g)^{abc}$. If we let $ab=\sum v_k$ and $c=s_0$ (this is possible because $v_k, k\in\lbrace 1,2,\cdots,N\rbrace $ and $s_0$ are all randomly chosen from $\mathbb{Z}_p$), we have $Z=e(g,g)^{abc}=(e(g,g)^{ab})^c=Y^{s_0}$. Therefore, $\mathbf{CT^{*}}$ is a valid ciphertext of the message $m_{\gamma}$. Otherwise, if $\mu=1$, $Z=e(g,g)^{z}$. Then, we have $E_0=m_{\gamma}\cdot e(g,g)^z$. Since $z\in\mathbb{Z}_p$ is a random element, $E_0$ is a random element in $\mathbb{G}_T$ from $\mathcal{A}$'s perspective, therefore $\mathbf{CT^{*}}$ contains no information about $m_{\gamma}$.
\\
\textsf{Phase 2} \hspace{4pt}
Repeat Phase 1 adaptively.
\\
\textsf{Guess} \hspace{4pt}
$\mathcal{A}$ submits a guess $\gamma'$ of $\gamma$. If $\gamma'=\gamma$, $sim$ outputs $\mu'=0$, indicating that it was given a valid DBDH-tuple $(g,A,S,Z)$, otherwise it outputs $\mu'=1$, indicating that he was given a random 5-element tuple $(g,A,B,C,Z)$.

As shown in the construction of the game, the simulator $sim$ computes the public parameter and the private key in the same way as our scheme. When $\mu=1$, the adversary $\mathcal{A}$ learns no information about $\gamma$, so we have \mbox{Pr$[\gamma\neq \gamma'|\mu=1]=\frac{1}{2}$}. Since the challenger guesses $\mu'=1$ when $\gamma=\gamma'$, we have \mbox{Pr$[\mu'=\mu|\mu=1]=\frac{1}{2}$}. If $\mu=0$, the adversary $\mathcal{A}$ gets a valid ciphertext of $m_\gamma$. $\mathcal{A}$'s advantage in this situation is $\epsilon$ by definition, so we have \mbox{Pr$[\gamma=\gamma'|\mu=0]=\frac{1}{2}+\epsilon$}. Since the challenger guesses $\mu'=0$ when $\gamma=\gamma'$, we have Pr$[\mu'=\mu|\mu=0]=\frac{1}{2}+\epsilon$. The overall advantage in this DBDH game is \mbox{$\frac{1}{2}$Pr$[\mu'=\mu|\mu=0]+\frac{1}{2}$Pr$[\mu'=\mu|\mu=1]-\frac{1}{2}$}
\\$=\frac{1}{2}\cdot(\frac{1}{2}+\epsilon)+\frac{1}{2}\cdot\frac{1}{2}-\frac{1}{2}=\frac{\epsilon}{2}$.

To conclude, as proved above, the advantage for a polynomial-time adversary in the DBDH game is $\frac{\epsilon}{2}$ if the advantage for a polynomial-time adversary in our security model is $\epsilon$. Therefore, if an adversary can break our scheme in our security model, which indicates $\epsilon$ is a non-negligible advantage, a polynomial-time adversary's advantage, which is $\frac{\epsilon}{2}$, in solving the DBDH problem is also non-negligible.
\end{proof}

Since our scheme relies on the assumption that no probabilistic polynomial algorithm can solve the DBDH problem with non-negligible advantage, it can be deduced that no adversary can break our scheme in our security model.

\begin{figure*}[!tH]
  \begin{center}
    \subfigure[Setup time]{\label {fig:result-a}\includegraphics[scale=0.17]{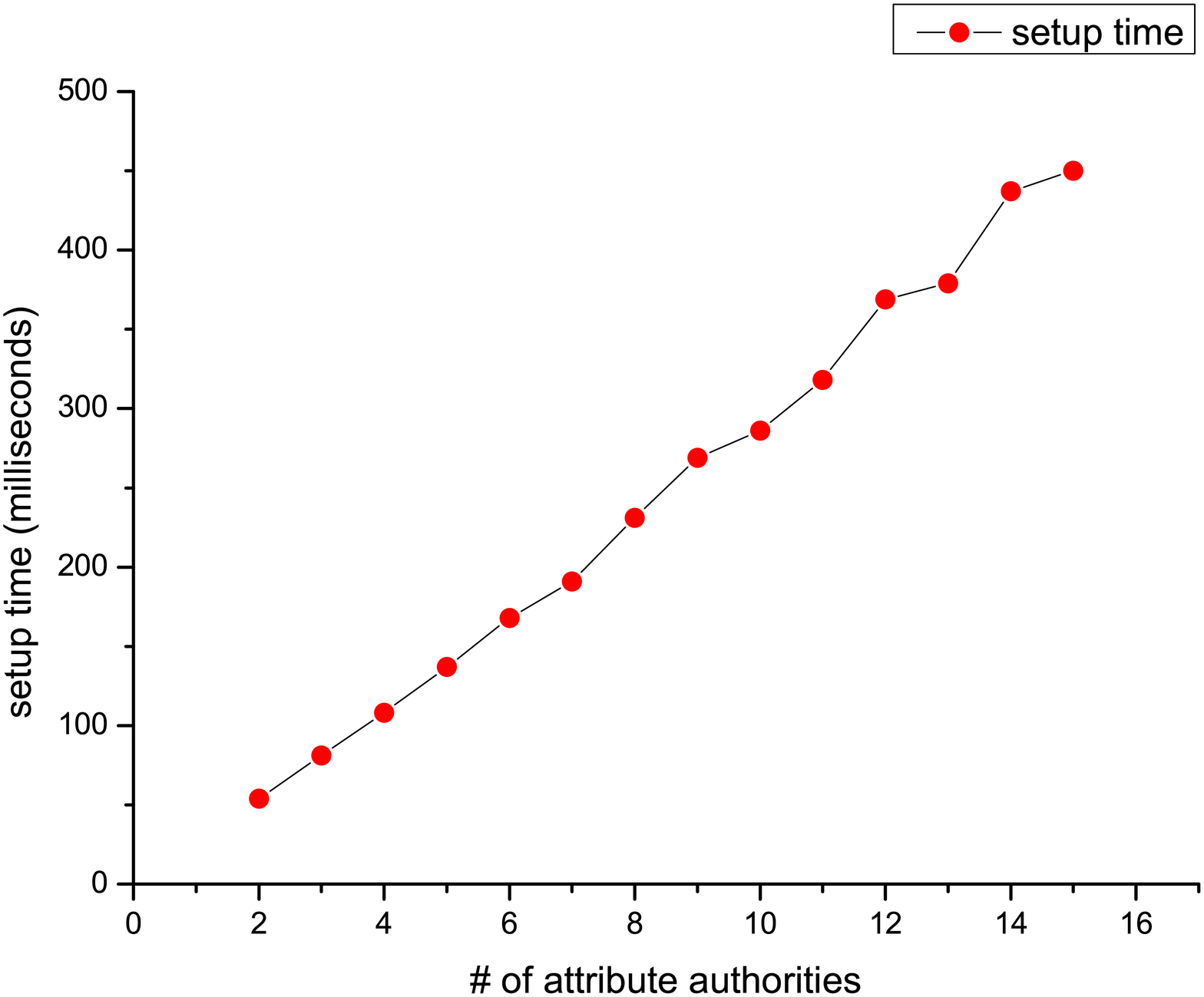}}
    \subfigure[Keygen time with different authorities number]{\label {fig:result-b}\includegraphics[scale=0.17]{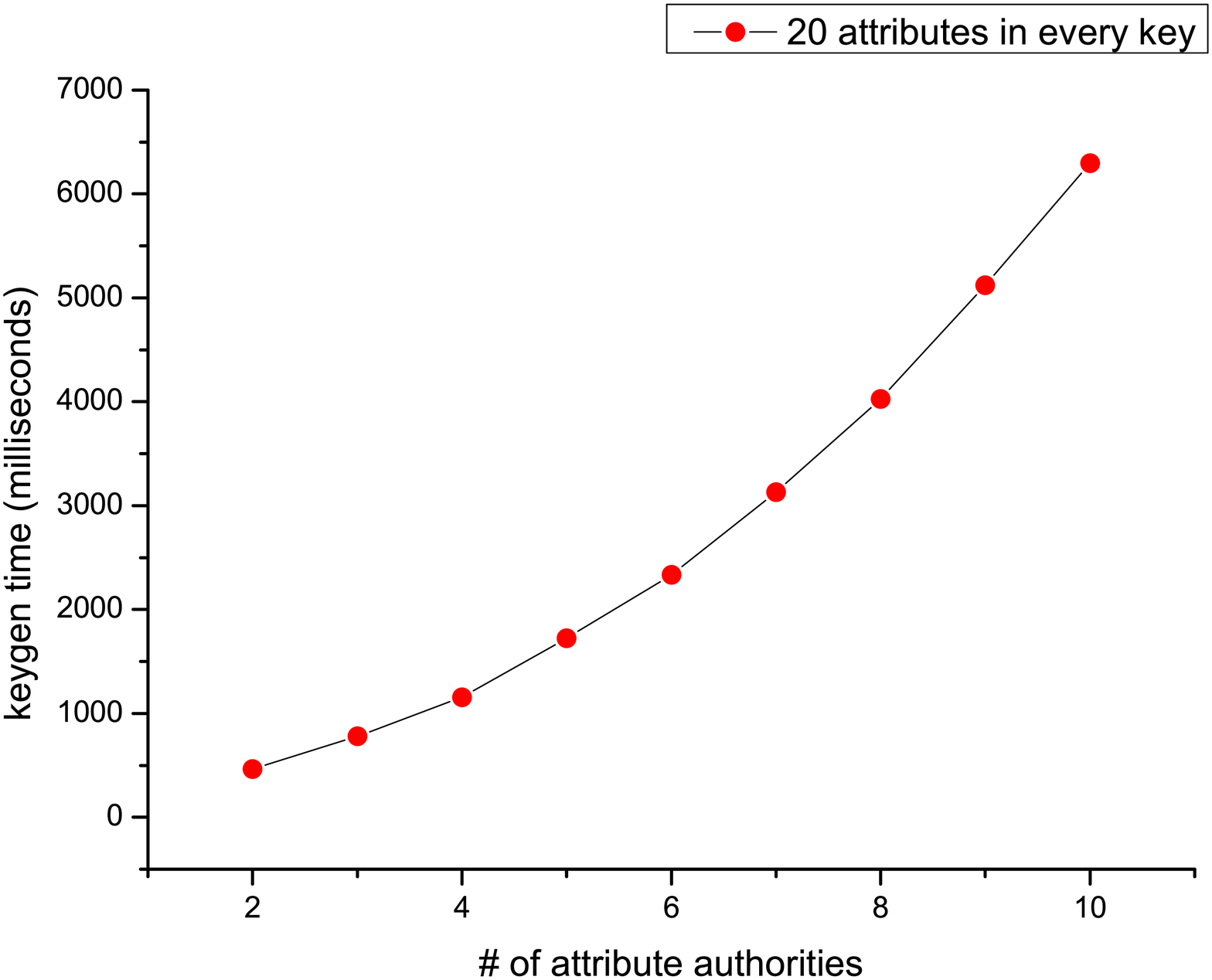}}
    \subfigure[Keygen time with different attributes number]{\label {fig:result-c}\includegraphics[scale=0.17]{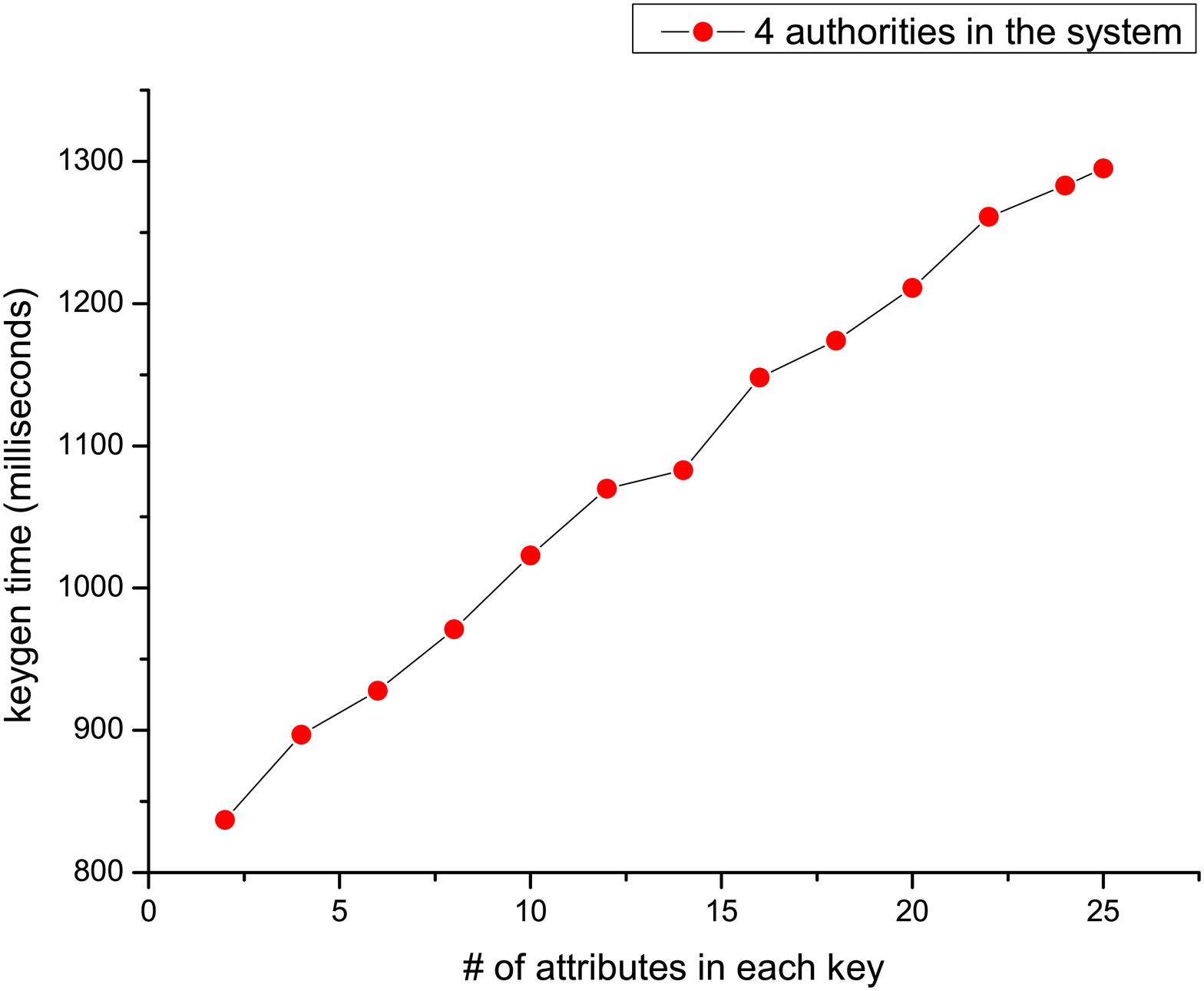}}
    \subfigure[Encryption and decryption time with different attributes number. File size is 100KB]{\label {fig:result-d}\includegraphics[scale=0.17]{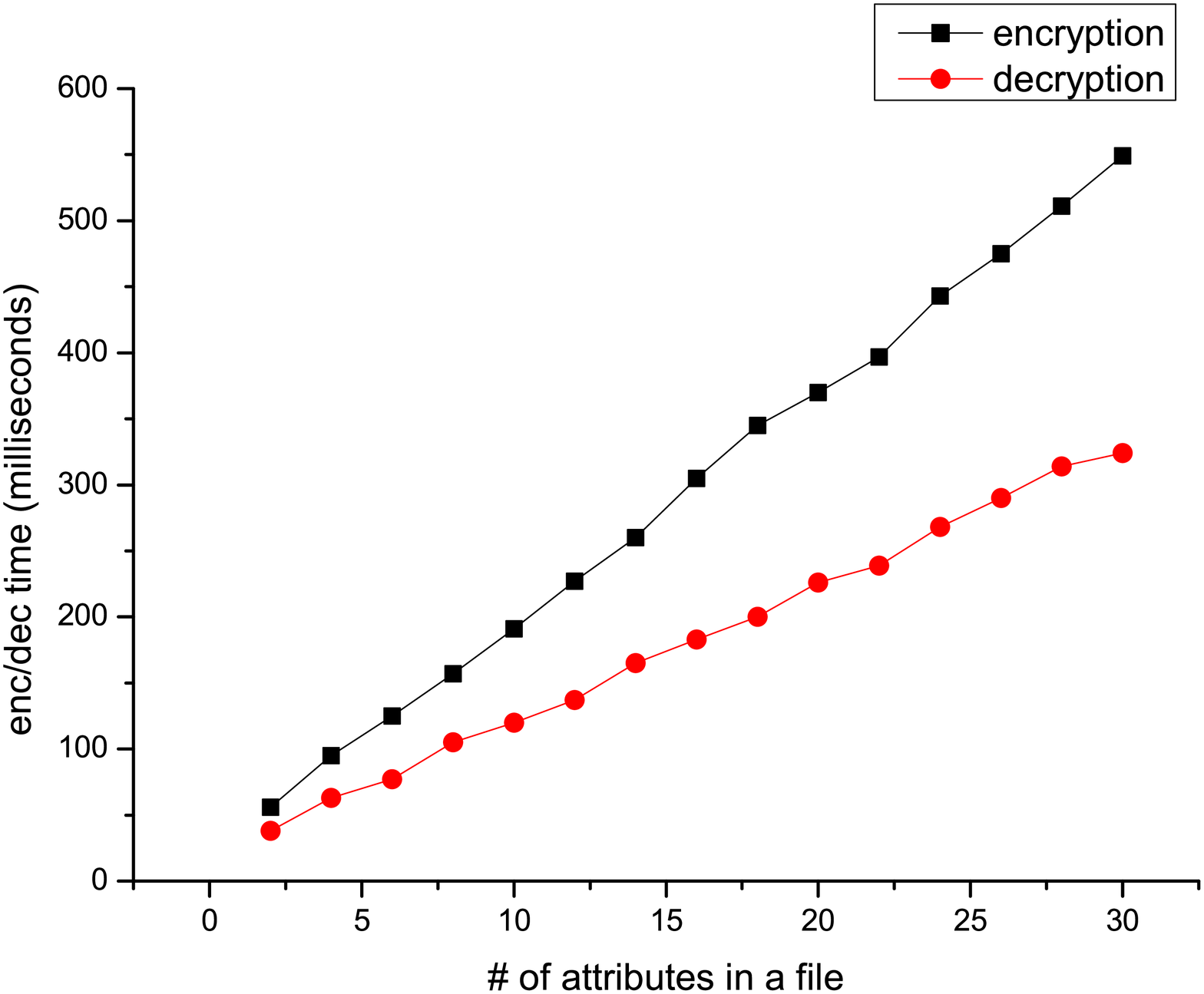}}
    \subfigure[Encryption and decryption time with different file size. 20 attributes in $T_0$]{\label {fig:result-e}\includegraphics[scale=0.17]{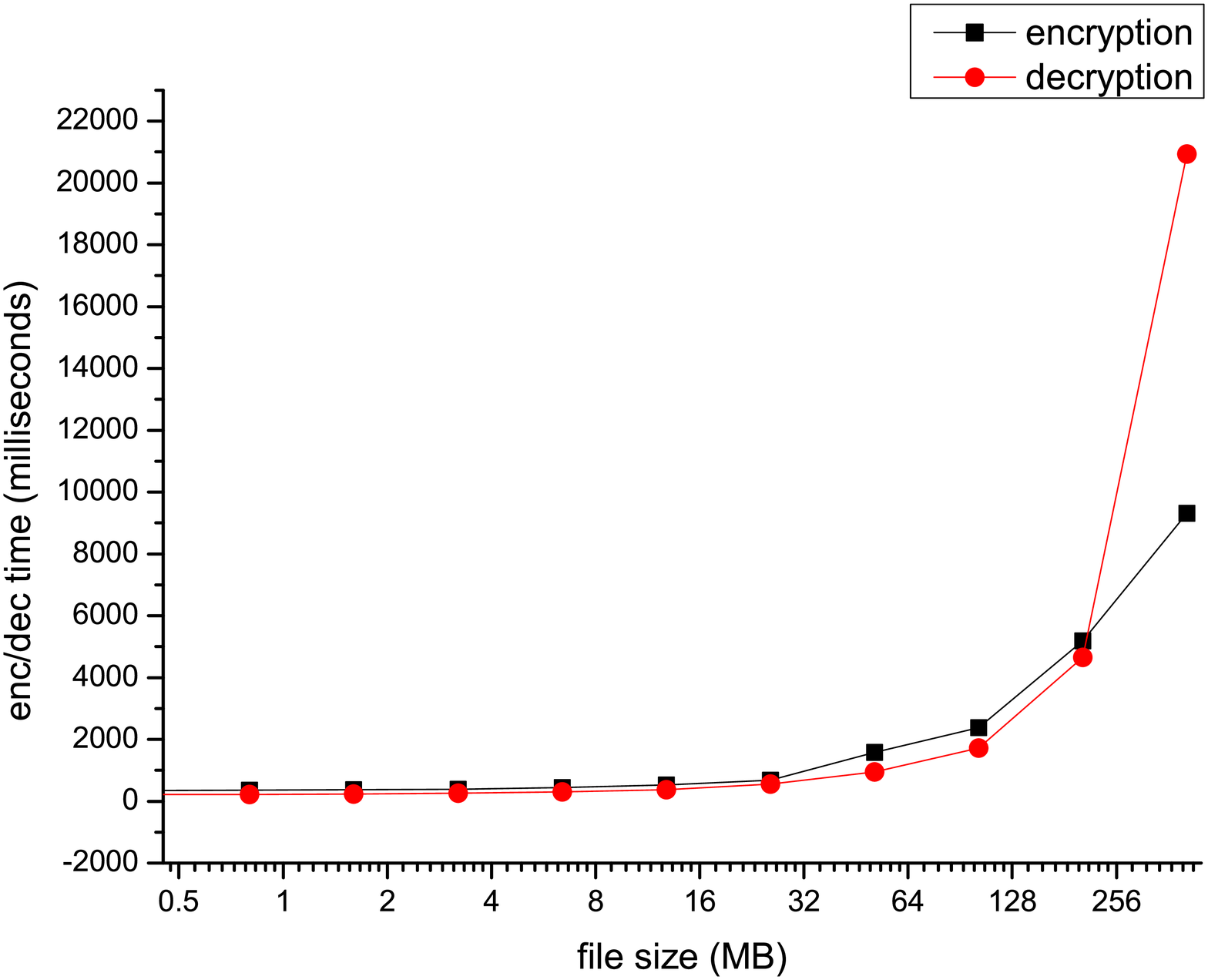}}
    \subfigure[Time to create a privilege tree and decrypt a verification parameter from it]{\label {fig:result-f}\includegraphics[scale=0.17]{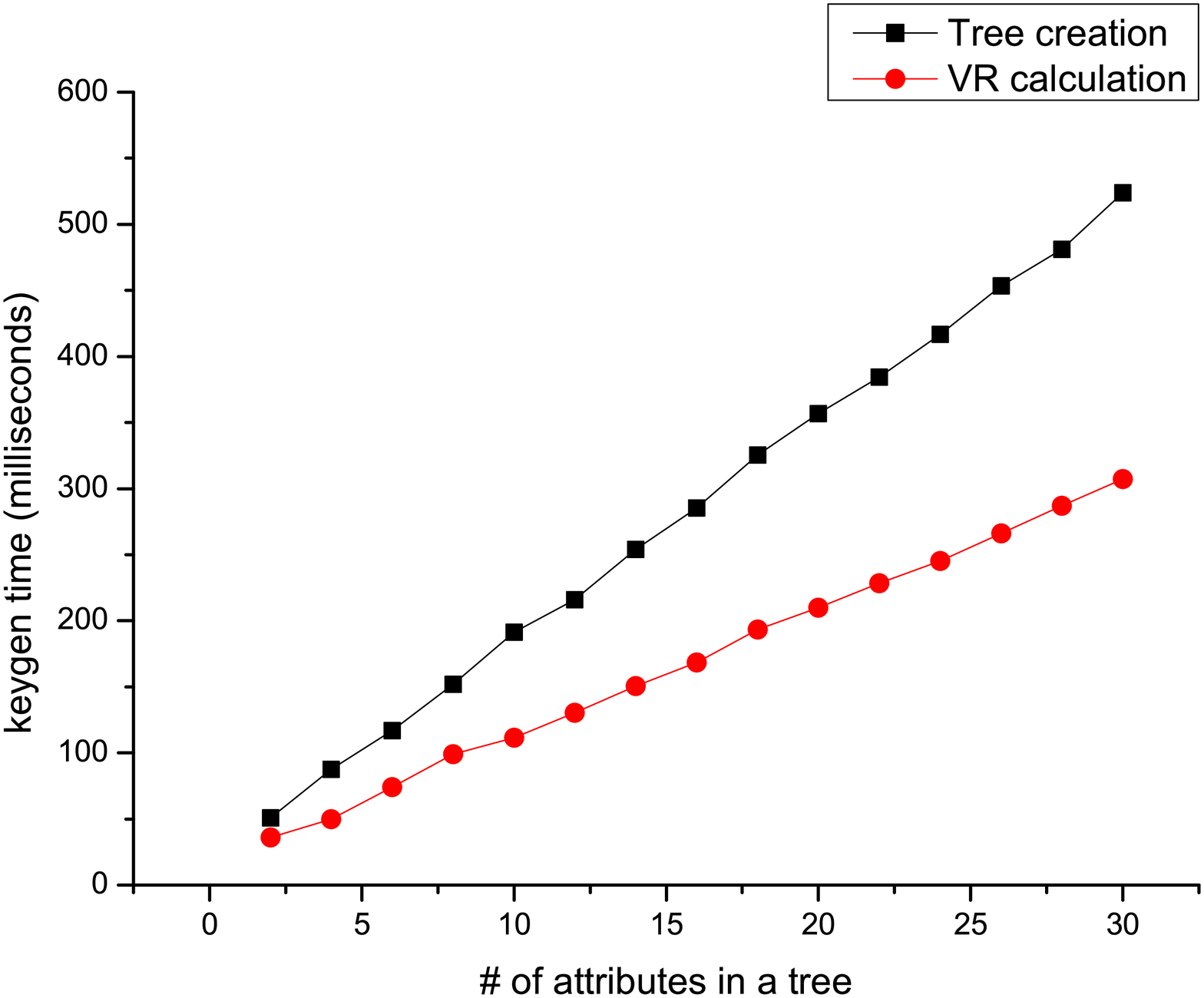}}
  \end{center}
  \vspace{-5pt}
  \caption{Experiment result on our implemented prototype system}
  \vspace{-8pt}
  \label{fig:result}
\end{figure*}

\section{Performance Analysis}
In this section, we denote $N$ as the number of attribute authorities, $I$ as the size of the entire attribute set and $X$ as the number of nodes in a tree $T_p$.

\subsection{Setup}
When the system is setup, $\prod{Y_k}$ is computed by any one of the authorities and sent to others, whose complexity is $O(N)$. Then, secret parameters $x_k$'s are calculated within the clusters. The complexity of that calculation is $O(C^2\cdot\frac{N}{C})=O(C\cdot N)$, but $C$ is a constant number, so $O(C\cdot N)=O(N)$. Therefore, the total complexity is $O(N)$. However, since we have $N$ authorities per system, the complexity per authority is $O(1)$. 
\subsection{Key Generation}
In the Attribute Key Generation, $g^{\sum v_j}$ is computed by $N$ authorities, and $D_i=H(att(i))^{r_i}\cdot{g^{\sum v_j}}$ is computed for $I$ times by one attribute authority. Therefore, the total complexity of Attribute Key Generation is $O(N^2+I\cdot N)$. In the Aggregation of Two Keys, a user aggregates the $I$ components, thus the computation complexity of this operation is $O(I)$. So, the complexity per authority is $O(N+I)$.
\subsection{Encryption}
At every non-leaf node, a polynomial is chosen and $k_x-1$ numbers are to be found to determine the polynomial, where $k_x$ is the threshold value. Therefore, denoting the average threshold value to be $K$,the computation complexity of this process is $O(X\cdot K)$.
\subsection{Decryption}
$DecryptNode$ is a recursive algorithm, and it is executed exactly once at every nodes in a Breadth-First-Search manner, therefore the computation complexity of this process is $O(X)$.
\subsection{User Revocation}
This operation has the same complexity as the addition of Decryption and Encryption, thus its complexity is $O(X\cdot K)$.

\begin{table}[!t]
\centering 
\vspace{-5pt}
\caption{Complexity Comparison (per authority)}
\vspace{-5pt}
\label{table:complexity}
\centering
\begin{tabular}{r|c|c|c}
\hline \hline
Process    & Yu \emph{et al.} \cite{kp-abe-cloud}& Chase \emph{et al.} \cite{multi2} & Ours\\
\hline
Setup & $O(I)$ & $O(1)$ &$O(1)$\\
Key Generation & $O(X)$ & $ O(N+I)$  & $O(N+I)$\\
Encryption & $O(I)$ & $O(I)$ & $O(X\cdot K)$\\
Decryption & $O(\max(X,I))$ & $O(N\cdot I)$ & $O(X)$\\
User Revocation &  $O(I)$ &  & $O(X\cdot K)$ \\
\hline \hline
\end{tabular}
\end{table}
\vspace{-18pt}

\section{Implementation}

In this section, we give the experimental result of our scheme, which is conducted on the prototype of our scheme. To the best of our knowledge, this is the first implementation of a multi-authority attribute based encryption scheme. Our prototype system provides five command line tools.

\noindent \textsf{anonyabe-setup} : Jointly generates a public key and $N$ master keys.

\noindent \textsf{anonyabe-keygen} : Generates a part of private key for the attribute set it is responsible for.

\noindent \textsf{anonyabe-enc} : Encrypts a file under $r$ privilege trees.

\noindent \textsf{anonyabe-dec} : Decrypts a file if possible.

\noindent \textsf{anonyabe-rec} : Re-encrypts a file under other privilege trees.

This toolkit is based on the CP-ABE toolkit \cite{cp-abe} which is in turn based on PBC library, and the whole system is implemented on a linux system with Intel i7 $2^{nd}$ Gen @ 2.7GHz and 2GB RAM.

Figure.\ref{fig:result} shows the computation overhead incurred in the core algorithms \textsf{Setup}, \textsf{Key Generation}, \textsf{Encryption} and \textsf{Decryption}  under various conditions.

Figure.\ref{fig:result-a} shows the system-wide setup time with different number of attribute authorities. Figure.\ref{fig:result-b} shows the total key generation time (system-wide) with different number of authorities, and the number of attributes is fixed to 20. Figure.\ref{fig:result-c} shows the key generation time with different number of attributes in each key, and the number of authorities is fixed to 4. Figure\ref{fig:result-d} shows the encryption and decryption time with different number of attributes in $T_0$, and we set only one privilege for file access to measure the most frequent operation, file access. Figure.\ref{fig:result-e} shows the encryption and decryption time with different file sizes, where the number of attributes in $T_0$ is fixed to 20. Figure.\ref{fig:result-f} shows the time to create a privilege tree and calculate a verification parameter $Y^{s_j}$ from it. Obviously, the total time needed to create one $\mathbf{VR}$ is approximately equal to $r\cdot t$, where $r$ is the number of total privileges and $t$ is the time for creating one tree.

The \textsf{Re-encryption} is omitted because it is barely a composition of \textsf{Decryption} and \textsf{Encryption}. Interestingly, in a series of the experiment, the run time of encryption and decryption was independent of the tree structure. That is, no matter how complicated the tree is, the computation complexity of encryption and decryption depends only on the number of nodes in the tree, which coincides with the performance analysis table in the previous section. Results of other algorithms are just as we expected. The graphs generally followed the growth rate showed in the performance analysis table above.

\section{Conclusion}

This paper proposed an anonymous attribute-based privilege control scheme \emph{AnonyControl} to
address the user privacy problem in a cloud storage server. Using multiple authorities in the cloud
computing system, our proposed scheme achieves not only fine-grained privilege control, but also
anonymity while conducting privilege control based on users' identity information. More importantly, our system can tolerate
 up to $N-2$ authority compromise, which is highly preferable especially in Internet-based cloud
computing environment. Furthermore, although the data contents are fully outsourced to Cloud Servers, the Cloud Servers cannot read the contents unless their private keys satisfy the privilege tree $T_0$.
We also conducted detailed security and performance analysis which shows that \emph{AnonyControl} is
both secure and efficient for cloud storage system.


\begin{thebibliography}{10}
\providecommand{\url}[1]{#1}
\csname url@samestyle\endcsname
\providecommand{\newblock}{\relax}
\providecommand{\bibinfo}[2]{#2}
\providecommand{\BIBentrySTDinterwordspacing}{\spaceskip=0pt\relax}
\providecommand{\BIBentryALTinterwordstretchfactor}{4}
\providecommand{\BIBentryALTinterwordspacing}{\spaceskip=\fontdimen2\font plus
\BIBentryALTinterwordstretchfactor\fontdimen3\font minus
  \fontdimen4\font\relax}
\providecommand{\BIBforeignlanguage}[2]{{%
\expandafter\ifx\csname l@#1\endcsname\relax
\typeout{** WARNING: IEEEtran.bst: No hyphenation pattern has been}%
\typeout{** loaded for the language `#1'. Using the pattern for}%
\typeout{** the default language instead.}%
\else
\language=\csname l@#1\endcsname
\fi
#2}}
\providecommand{\BIBdecl}{\relax}
\BIBdecl

\bibitem{shamir1985identity}
A.~Shamir, ``Identity-based cryptosystems and signature schemes,'' in
  \emph{Advances in cryptology},  Springer, 1985, pp. 47--53.

\bibitem{abe}
A.~Sahai and B.~Waters, ``Fuzzy identity-based encryption,'' \emph{Advances in
  Cryptology--EUROCRYPT 2005}, pp. 557--557, 2005.

\bibitem{kp-abe}
V.~Goyal, O.~Pandey, A.~Sahai, and B.~Waters, ``Attribute-based encryption for
  fine-grained access control of encrypted data,'' in \emph{Proceedings of the
  13th ACM conference on Computer and communications security}, 2006, pp. 89--98.

\bibitem{cp-abe}
J.~Bethencourt, A.~Sahai, and B.~Waters, ``Ciphertext-policy attribute-based
  encryption,'' in \emph{IEEE Symposium
  on Security and Privacy, 2007}, pp. 321--334.

\bibitem{multi1}
M.~Chase, ``Multi-authority attribute based encryption,'' \emph{Theory of
  Cryptography}, pp. 515--534, 2007.

\bibitem{multi2}
M.~Chase and S.~Chow, ``Improving privacy and security in multi-authority
  attribute-based encryption,'' in \emph{Proceedings of the 16th ACM conference
  on Computer and communications security}, 2009, pp. 121--130.

\bibitem{hibe}
J.~Horwitz and B.~Lynn, ``Toward hierarchical identity-based encryption,'' in
  \emph{Advances in Cryptology--EUROCRYPT 2002}, Springer, 2002, pp. 466--481.

\bibitem{efficient-revocation}
A.~Boldyreva, V.~Goyal, and V.~Kumar, ``Identity-based encryption with
  efficient revocation,'' in \emph{Proceedings of the 15th ACM conference on
  Computer and communications security}, 2008, pp. 417--426.

\bibitem{hidden-tree}
T.~Nishide, K.~Yoneyama, and K.~Ohta, ``Attribute-based encryption with
  partially hidden encryptor-specified access structures,'' in \emph{Applied
  Cryptography and Network Security},
  Springer, 2008, pp. 111--129.

\bibitem{ostrovsky2007abe}
R.~Ostrovsky, A.~Sahai, and B.~Waters, ``Attribute-based encryption with
  non-monotonic access structures,'' in \emph{Proceedings of the 14th ACM
  conference on Computer and communications security}, 2007, pp. 195--203.

\bibitem{lewko2011decentralizing}
A.~Lewko and B.~Waters, ``Decentralizing attribute-based encryption,''
  \emph{Advances in Cryptology--EUROCRYPT 2011}, pp. 568--588, 2011.

\bibitem{kp-abe-cloud}
S.~Yu, C.~Wang, K.~Ren, and W.~Lou, ``Achieving secure, scalable, and
  fine-grained data access control in cloud computing,'' in \emph{IEEE INFOCOM, 2010}, pp.
  1--9.

\bibitem{habe}
G.~Wang, Q.~Liu, and J.~Wu, ``Hierarchical attribute-based encryption for
  fine-grained access control in cloud storage services,'' in \emph{Proceedings
  of the 17th ACM conference on Computer and communications security}, 2010, pp. 735--737.

\bibitem{hasbe}
J.~Liu, Z.~Wan, and M.~Gu, ``Hierarchical attribute-set based encryption for
  scalable, flexible and fine-grained access control in cloud computing,''
  \emph{Information Security Practice and Experience}, pp. 98--107, 2011.

\bibitem{revoke-cloud}
S.~Yu, C.~Wang, K.~Ren, and W.~Lou, ``Attribute based data sharing with
  attribute revocation,'' in \emph{Proceedings of the 5th ACM Symposium on
  Information, Computer and Communications Security}, 2010, pp. 261--270.

\bibitem{hidden-credential}
A.~Kapadia, P.~Tsang, and S.~Smith, ``Attribute-based publishing with hidden
  credentials and hidden policies,'' \emph{NDSS--07}, pp. 179--192, 2007.

\bibitem{hidden-policy}
S.~Yu, K.~Ren, and W.~Lou, ``Attribute-based content distribution with hidden
  policy,'' in \emph{4th Workshop
  on Secure Network Protocols, 2008}, pp. 39--44.

\bibitem{jung2013data}
T.~Jung, X.~Mao, X.~Li, S.~Tang, W.~Gong, and L.~Zhang, ``Privacy-preserving
  data aggregation without secure channel: multivariate polynomial
  evaluation,'' in \emph{IEEE INFOCOM}, 2013.

\bibitem{li2013search}
X.~Li and T.~Jung, ``Search me if you can: privacy-preserving location query
  service,'' in \emph{IEEE INFOCOM}, 2013.

\bibitem{zhang2013verifiable}
L.~Zhang, X.~Li, Y.~Liu, and T.~Jung, ``Verifiable private multi-party
  computation: ranging and ranking,'' in \emph{IEEE INFOCOM Mini-Conference},
  2013.

\bibitem{zhang2012message}
L.~Zhang and X.~Li, ``Message in a sealed bottle: Privacy preserving friending
  in social networks,'' \emph{arXiv preprint arXiv:1207.7199}, 2012.

\bibitem{public-audit}
C.~Wang, Q.~Wang, K.~Ren, and W.~Lou, ``Privacy-preserving public auditing for
  data storage security in cloud computing,'' in \emph{IEEE INFOCOM, 2010}.
 pp.  1--9.

\bibitem{linear-outsourcing}
C.~Wang, K.~Ren, and J.~Wang, ``Secure and practical outsourcing of linear
  programming in cloud computing,'' in \emph{IEEE INFOCOM, 2011}, pp. 820--828.

\bibitem{rank-search}
C.~Wang, N.~Cao, J.~Li, K.~Ren, and W.~Lou, ``Secure ranked keyword search over
  encrypted cloud data,'' in \emph{IEEE ICDCS, 2010}, pp. 253--262.

\bibitem{liu2011rumor}
Y.~Liu, J.~Han, and J.~Wang, ``Rumor riding: anonymizing unstructured
  peer-to-peer systems,'' \emph{IEEE
  Transactions on Parallel and Distributed Systems}, vol.~22, no.~3, pp. 464--475, 2011.

\bibitem{shamir1979share}
A.~Shamir, ``How to share a secret,'' \emph{Communications of the ACM},
  vol.~22, no.~11, pp. 612--613, 1979.

\end{thebibliography}


\end{document}